\documentclass[notitlepage,english,aps,floats,onecolumn,showpacs,nofootinbib,floatfix]{revtex4-1}

\usepackage{pslatex}
\usepackage[T1]{fontenc}
\usepackage[latin1]{inputenc}
\usepackage{graphicx}
\usepackage{epsfig}
\usepackage{longtable}
\usepackage{float}
\usepackage{calc}
\usepackage{ifthen}
\usepackage{amsmath}
\usepackage{hyperref}
\usepackage{amssymb}

\usepackage{color}

{
{
{
\newcommand{\bea}{\begin{eqnarray}}
\newcommand{\eea}{\end{eqnarray}}

\newcommand{\nc}{\newcommand}
\nc{\renc}{\renewcommand}
\nc{\eqs}[2]{\mbox{Eqs.~(\ref{#1},\,\ref{#2})}}
\nc{\eq}[1]{\mbox{Eq.~(\ref{#1})}}
\nc{\figs}[2]{\mbox{Figs.~(\ref{#1},\,\ref{#2})}}
\nc{\fig}[1]{\mbox{Fig~.(\ref{#1})}}
\nc{\be}[1]{\begin{equation} \mbox{$\label{#1}$}}
\nc{\ee}{\vspace{0.1cm}\end{equation}}

\newcommand{\bean}{\begin{eqnarray*}}
\newcommand{\eean}{\end{eqnarray*}}

\def\eV{{\rm \ eV}}
\def\GeV{{\rm \ GeV}}

\def\lae{\;^{<}_{\sim} \;} \def\gae{\; ^{>}_{\sim} \;}

\begin{document}

\title{Leptogenesis via Inflaton Mass Terms in Non-Minimally Coupled Inflation }

\author{Kit Lloyd-Stubbs  and John McDonald }
\email{a.lloyd-stubbs@lancaster.ac.uk}
\email{j.mcdonald@lancaster.ac.uk}
\affiliation{Dept. of Physics,  
Lancaster University, Lancaster LA1 4YB, UK}

\begin{abstract}

We consider a model of baryogenesis based on adding lepton number-violating quadratic mass terms to the inflaton potential of a non-minimally coupled inflation model. The $L$-violating mass terms generate a lepton asymmetry in a complex inflaton field via the mass term Affleck-Dine mechanism, which is transferred to the Standard Model (SM) sector when the inflaton decays to right-handed (RH) neutrinos. The model is minimal in that it requires only the SM sector, RH neutrinos, and a non-minimally coupled inflaton sector.   We find that baryon isocurvature fluctuations can be observable in metric inflation but are negligible in Palatini inflation. The model is compatible with reheating temperatures that may be detectable in the observable primordial gravitational waves predicted by metric inflation.

\end{abstract}
 
\maketitle

\section{Introduction}

Baryogenesis is an essential feature of a complete model of particle cosmology. In the absence of a fundamental theory of particle physics, a minimal model-building approach seems most likely to lead to the correct solution,  by virtue of its fewer assumptions. A minimal model of particle physics and cosmology consists of the Standard Model sector, neutrino masses via right-handed (RH) neutrinos, a dark matter candidate, and an inflaton sector. In this case there is a limited range of possibilities for generating the baryon asymmetry. The most commonly considered are electroweak baryogenesis \cite{ew} and leptogenesis via out-of-equilibrium decays of RH neutrinos \cite{lept,lept2,lept3}. 

Another possibility is the generation of the asymmetry via the decay of the inflaton. This could occur via out-of-equilibrium decay of the inflaton if it has the required CP- and B-violating decay modes \cite{infb,infb2}. Alternatively, the asymmetry could first be generated in the inflaton field itself via the Affleck-Dine (AD) \cite{ad} mechanism   \cite{adc0,adc1,adc1a,adc2,adc3,cline,clineL,kls1,kawa}. This is arguably the simplest mechanism for generating the baryon asymmetry, requiring only $(B-L)$-violating terms in the potential of a complex scalar inflaton and a decay mode to transfer the asymmetry to the SM sector.   

In a previous paper \cite{kls1} we discussed a new implementation of AD inflation, in which the asymmetry is generated at late times by $(B-L)$-violating quadratic potential terms (mass terms) during the rapid coherent oscillations of a complex inflaton\footnote{The AD mechanism based on mass terms was previously considered in a different context in \cite{gorb}. This analysis does not include the case of final asymmetry generation from averaging over a rapidly oscillating AD asymmetry.}. This results in an oscillating asymmetry in the inflaton condensate, with mean value equal to zero. Nevertheless, when the asymmetry is transferred to the SM sector, the net asymmetry is non-zero.

Inflation models are constrained by present bounds on the range of scalar spectral index values and the upper bound on the tensor-to-scalar ratio. One class of models that provides a good fit to observations are the non-minimally coupled inflation models  \cite{bbs}. There are two favoured implementations of this model, based on either the metric formalism of General Relativity (GR), as in conventional Higgs Inflation \cite{ks}, or the Palatini formalism \cite{palatiniHI}. 
Such models can be driven by a renormalisable $ \lambda \phi^4$ inflaton potential with a value for $\lambda$ that is not very small compared to 1. Non-minimally coupled inflation therefore provides a good candidate for the inflaton sector of a minimal model of particle cosmology. 

In our previous paper \cite{kls1} we discussed the general idea of baryogenesis via inflaton mass terms. However, we did not consider our results in the context of a specific inflation model. In this paper we will present a minimal model based on non-minimally coupled inflation.  Inflaton mass term AD baryogenesis and leptogenesis has also recently been applied in the context of models which relate inflation and baryon asymmetry generation to  neutrino-antineutrino oscillations \cite{mo1}; to neutrino masses combined with pseudo-Goldstone dark matter \cite{mo2}, WIMP dark matter \cite{mo3} and axion dark matter \cite{mo4}; and to cogenesis of the baryon asymmetry and dark matter density \cite{okada}.

  The minimal model we consider consists of the SM sector extended by RH neutrinos plus a complex inflaton sector, with a global $U(1)_{L}$ symmetry. To this we add $U(1)_{L}$-breaking mass terms for the inflaton and the RH neutrinos.  The baryon asymmetry is generated by creating a lepton asymmetry in the inflaton field via the mass term AD mechanism, which is subsequently transferred to the SM sector via inflaton decays to RH neutrinos. 

    The paper is organised as follows. In Section 2 we introduce our leptogenesis model. In Section 3 we review the essential results of non-minimally coupled inflation in the metric and Palatini formalisms. 
 In Section 4 we review and elaborate upon the general results of \cite{kls1} for the mass term AD mechanism and apply the results to our leptogenesis model. In Section 5 we discuss condensate decay, reheating and the conditions to ensure no asymmetry washout. We also discuss the consistency of the assumptions made in the derivation of the analytical results. In Section 6 we discuss the possibility of generating observable baryon isocurvature perturbations. 
In Section 7 we discuss the conditions under which the quadratic $U(1)_{L}$-breaking inflaton potential terms dominate the quartic $U(1)_{L}$-breaking terms. 
In Section 8 we present our conclusions. 

\section{A Minimal Model for Affleck-Dine Leptogenesis via Inflaton Mass Terms}

\subsection{The Observable Sector}

A complex inflaton $\Phi$ has a general renormalisable potential given by\footnote{We do not include a linear term as this can always be eliminated by a redefiniton of $\Phi$ and the $A$, $B$ and  $C$ terms.}
\be{e1}  V(\Phi) = m_{\phi}^{2} |\Phi|^{2}  + \lambda_{\Phi} |\Phi|^{4} - (A \Phi^2 + {\rm h.\,c.}) - (B \Phi^{3} + {\rm h.\,c.}) - (C \Phi^{4} + {\rm h.\,c.})  ~. \ee
In the limit $A = B = C = 0$ this has a global $U(1)$ symmetry, $\Phi \rightarrow e^{i \alpha} \Phi$, which corresponds to lepton number in the model we are considering.  The $A$, $B$ and $C$ terms are $L$-violating terms that can be used to generate a lepton asymmetry in the inflaton condensate via the AD mechanism. For simplicity, we will assume that there is an unbroken discrete symmetry of the potential, $\Phi \leftrightarrow -\Phi$, so that $B = 0$. Generation of the baryon asymmetry in non-minimally coupled inflation via the quartic $C$-term  has previously been considered in \cite{cline} and \cite{clineL}. Here we will focus on generation via the $A$-term \cite{kls1}, which has a quite different dynamics.

In order to transfer the asymmetry to the SM sector, we couple the inflaton to the RH neutrinos via  
\be{e2}   {\cal L}_{int} =   - y_{\Phi} \Phi \overline{N_{R}^{c}} N_{R} - h_{\nu}\overline{N_{R}^{c}} HL -  m_{N} \overline{N_{R}^{c}} N_{R} + {\rm h.\,c.}   ~,\ee
where $\Phi$ has lepton number $-2$. We have also included the Yukawa coupling $h_{\nu}$ of the RH neutrino to the Higgs $H$ and the lepton $L$ doublets, and an $L$-violating mass for the RH neutrinos. These result in a Type-I neutrino mass matrix once the Higgs expectation value $v$ is introduced. For the case of a single generation, which we will use as a representative example, this results in a heavy eigenvalue with mass approximately equal to $m_{N}$ and a light eigenvalue $m_{\nu} \approx m_{D}^{2}/m_{N}$ corresponding to the mass of the observed neutrinos, where $m_{D} = h_{\nu} v/\sqrt{2}$.
$h_{\nu}$ is then related to the observed neutrino masses by 
\be{e4}  h_{\nu} = \frac{\sqrt{2}}{v} \left( m_{\nu} m_{N} \right)^{1/2}    ~,\ee
where $v = 246 \GeV$ is the Higgs vacuum expectation value.  

In addition to \eq{e2}, there can be a natural L-conserving portal coupling to the Higgs doublet of the form  $\lambda_{\Phi H} |\Phi|^{2} |H|^{2} $. Therefore the process of reheating could be due to either perturbative decay of the $\Phi$ condensate scalars to RH neutrinos or parametric resonance to Higgs bosons pairs. Other possibilities are decay or annihilation of the inflaton via the non-minimal coupling or graviton exchange.  Here we will focus on reheating via perturbative decay to RH neutrinos. We will comment on the conditions for perturbative decay to RH neutrinos to dominate reheating and on how the results could change if reheating was dominated by the portal coupling or other processes.

\subsection{The Inflation Sector}  

The renormalisable potential \eq{e1} is naturally compatible with non-minimally coupled inflation. The action of the inflaton sector is 
\be{i1}   S = \int d^{4} x \sqrt{-g} \left[ \left( 1 + \frac{2 \xi  |\Phi|^{2}}{M_{Pl}^{2}} \right) \frac{M_{Pl}^{2} R}{2}   - \partial_{\mu} \Phi^{\dagger} \partial^{\mu} \Phi  - V(\Phi) \right]  ~, \ee
with signature $(-, +, +, +)$. In general, inflation will occur along a random value of $\theta$, where $\Phi = \phi e^{i \theta}/\sqrt{2}$. During inflation we will consider the $A$-terms and $C$-terms to be negligible, in which case the action can be written as  
\be{i2}   S = \int d^{4} x \sqrt{-g} \left[ \left( 1 + \frac{ \xi \phi^{2}}{M_{Pl}^{2}} \right) \frac{M_{Pl}^{2} R}{2}   - \frac{1}{2} \partial_{\mu} \phi \partial^{\mu} \phi  - V(\phi) \right]  ~,\ee
where 
\be{i2a}  V(\phi) = \frac{m_{\Phi}^{2} \phi^{2}}{2}
 +   \frac{\lambda_{\Phi} \phi^4}{4}     ~.\ee
Inflation is analysed in the Einstein frame, with action  
\be{i4}   S_{E} = \int d^{4} x \sqrt{-\tilde{g}} \left[ \frac{M_{Pl}^{2} \tilde{R}}{2}   - \frac{1}{2} \left( \Omega^{2} +  \frac{6 s \xi^{2} \phi^{2}}{M_{Pl}^{2}}\right) 
\frac{\partial_{\mu} \phi \partial^{\mu} \phi}{\Omega^{4}}   - V_{E}(\phi) \right]  ~,\ee
where $s = 1$ for the metric formalism and $s = 0$ for the Palatini formalism, and the Einstein frame potential is 
\be{i4a}   V_{E}(\phi)  = \frac{V(\phi)}{\Omega^{4}}   ~.\ee  
The contraction of indices and $\tilde{R}$ are now defined in terms of $\tilde{g}_{\mu\;\nu}$, where $\tilde{g}_{\mu\;\nu}  = \Omega^{2} g_{\mu\;\nu}$ and
\be{i3} \Omega^{2} =  1 + \frac{ \xi \phi^{2}}{M_{Pl}^{2}} ~.\ee 

\section{Non-Minimally Coupled Inflation in the metric and Palatini formalisms} 

In this section we summarise the key results of non-minimally coupled inflation that are relevant to the leptogenesis model. 

\subsection{Inflation Observables} 

 The amplitude of the power spectrum is 
\be{i8} A_{s} = \frac{\lambda_{\Phi} N^{2}}{72 \pi^{2} \xi^{2} } \;\;\;\; {\rm (Metric)}  \;\;\;\;;\;\;\;\; A_{s} = \frac{\lambda_{\Phi} N^{2}}{12 \pi^{2} \xi } \;\;\;\; {\rm (Palatini)}   ~,\ee
where $N$ is the number of e-foldings corresponding to the pivot scale.
The scalar spectral index is the same for both metric and Palatini
\be{i9} n_{s} \approx 1 - \frac{2}{N}  ~.\ee
The tensor-to-scalar ratio is 
\be{i14} r = \frac{12}{N^2}  \;\;\;\; {\rm (Metric)} \;\;\;\;;\;\;\;\; r = \frac{2}{\xi N^2}  \;\;\;\; {\rm (Palatini)} ~.\ee

\subsection{The non-minimal coupling}  

From the observed amplitude of the power spectrum, $A_{s} = 2.1 \times 10^{-9}$, it follows from \eq{i8}  that: 
\be{i16} \xi = 820 N \sqrt{\lambda_{\Phi}} \;\;\;\; {\rm (Metric)}   ~\ee
and 
\be{i17} \xi = 4.0 \times 10^{6} N^{2} \lambda_{\Phi} \;\;\;\; {\rm (Palatini)}    ~.\ee
The corresponding values for $N = 55$ are $\xi = 4.5 \times 10^{4} \sqrt{\lambda_{\Phi}}$ for the metric case and $\xi = 1.2 \times 10^{10} \lambda_{\Phi}$ for the Palatini case. 
 
Using these, \eq{i14} with $N = 55$ gives for the tensor-to-scalar ratio $r \approx 0.004$ in the metric model and $r \approx 5.5 \times 10^{-14}/\lambda_{\Phi}$ for the Palatini model.

\subsection{$\phi$ for which non-minimally coupled dynamics are negligible}

  The values of $\phi$ at which the Einstein and Jordan frame actions become completely equivalent are
\be{i18} \phi < \phi_{c} = \frac{M_{Pl}}{\sqrt{6} \xi} \;\;\;\; {\rm (Metric) }   ~\ee 
and 
\be{i19} \phi < \phi_{c} =\frac{M_{Pl}}{\sqrt{\xi}}  \;\;\;\; {\rm (Palatini)}    ~.\ee

\section{Affleck-Dine via inflaton mass terms}

In \cite{kls1} we introduced AD baryogenesis via quadratic inflaton potential terms. In this section we will review and expand upon the analytical results of \cite{kls1}. In particular, we will compare the analytic predictions for the baryon/lepton number-to-entropy ratio with the results of a complete numerical calculation based on solving the field equations, where we will show almost perfect agreement with the analytical results.  We will then apply our analytical results to the leptogenesis model.

\subsection{Analytical Lepton Asymmetry}

In terms of $\Phi = (\phi_{1} + i \phi_{2})/\sqrt{2}$, the inflaton potential is
\be{e4v} V(\Phi) = \frac{1}{2} (m_{\Phi}^{2} - 2A) \phi_{1}^{2} 
+ \frac{1}{2} (m_{\Phi}^{2} + 2A) \phi_{2}^{2} + \frac{\lambda_{\Phi}}{4} (\phi_{1}^{2} + \phi_{2}^{2})^{2}  ~,\ee
where without loss of generality we can define $A$ to be real and positive by a $U(1)_{L}$ rotation of $\Phi$. The field equations, including the decay terms, are
\be{e5a} \ddot{\phi}_{1} + 3 H \dot{\phi}_{1} 
+ \Gamma_{\Phi} \dot{\phi_{1}}  = -  m_{1}^{2} \phi_{1} - \lambda_{\Phi}(\phi_{1}^{2} + \phi_{2}^{2}) \phi_{1} ~\ee
and
\be{e5b} \ddot{\phi}_{2} + 3 H \dot{\phi}_{2} + \Gamma_{\Phi} \dot{\phi}_{2} = -  m_{2}^{2} \phi_{2} - \lambda_{\Phi}(\phi_{1}^{2} + \phi_{2}^{2}) \phi_{2} ~,\ee
where 
\be{e6a} m_{1}^{2} = m_{\Phi}^{2} - 2A  \;\;\;,\;\;\;\;\; m_{2}^{2} = m_{\Phi}^{2} + 2A   ~,\ee
and $\Gamma_{\Phi}$ is the inflaton decay rate.
In the limit $\lambda_{\Phi} \rightarrow 0$ the equations for $\phi_{1}$ and $\phi_{2}$ are decoupled from each other, with coherently oscillating solutions for $\phi_{1}$ and $\phi_{2}$ with different angular frequencies $m_{1}$ and $m_{2}$. 

As long as the $A$-terms do not become dynamically significant until after the coherent oscillations are dominated by the $|\Phi|^{2}$ term in the potential, and neglecting the decay terms for now, the solution for the coherently oscillating field is accurately described by considering only the mass terms in the potential,
\be{ex2}  V(\Phi) = m_{\Phi}^{2} |\Phi|^{2} - (A \Phi^{2} + {\rm h.c.} )    \;\;\;;\;\;\;\;\; \phi < \phi_{*}   ~,\ee 
where $\phi_{*} = m_{\Phi}/\sqrt{\lambda_{\Phi}}$ is the value of $\phi$ (with $\Phi = \phi e^{i \theta}/\sqrt{2}$) below which the field equations become dominated by the $|\Phi|^2$ term. 
The field at late times, $t > t_{*}$, where $\phi = \phi_{*}$ at $t = t_{*}$,  is then 
\be{e7} \phi_{1} = \phi_{1,\;*} \left(\frac{a_{*}}{a}\right)^{3/2} \cos(m_{1} (t - t_{*}))  \;\;;\;\; \phi_{2} = \phi_{2,\;*} \left(\frac{a_{*}}{a}\right)^{3/2} \cos(m_{2} (t - t_{*} ))  ~,\ee    
assuming that the asymmetry at $t \leq t_{*}$ is zero. The lepton asymmetry in the condensate (with $L(\Phi) = -2$) is given by 
\be{e9} n_{L}(t) = i L(\Phi) \left(\Phi^{\dagger}  \dot{\Phi} -  \dot{\Phi}^{\dagger} \Phi \right)   = -2 \left(\dot{\phi}_{1} \phi_{2} - \dot{\phi}_{2} \phi_{1} \right) ~.\ee
Assuming that $2A \ll m_{\Phi}^{2}$, and neglecting inflaton decay for the moment, the lepton asymmetry to leading order in $A/m_{\Phi}^{2}$ is 
 \be{e11} n_{L}(t) = -2 \phi_{1,\;*} \phi_{2,\;*}  \left(\frac{a_{*}}{a}\right)^{3} \left[ m_{\Phi} \sin\left( \frac{2 A (t - t_{*}) }{m_{\Phi}} \right) + \frac{A}{m_{\Phi}} \sin \left(2 m_{\Phi} (t - t_{*})\right) \right]   ~.\ee
On averaging over the rapid coherent oscillations of $\phi$, the second term in \eq{e11} averages to zero and the condensate lepton asymmetry at $t > t_{*}$ is  
\be{e12}  n_{L}^{0}(t) = -2 \phi_{1,\;*} \phi_{2,\;*}  \left(\frac{a_{*}}{a}\right)^{3} m_{\Phi} \sin\left( \frac{2 A (t - t_{*}) }{m_{\Phi}} \right)  ~,\ee 
where the superscript $0$ denotes the absence of decays. Therefore the asymmetry in the condensate will oscillate about zero with period $T_{asy}$ given by
\be{e14} T_{asy} = \frac{\pi m_{\Phi}}{A}   ~.\ee
Defining the comoving lepton asymmetry in the inflaton field by $n_{L\;c}(t) \equiv (a(t)/a_{*})^{3} n_{L}(t)$, where $a_{*}$ 
is the scale factor at $t_{*}$ ($n_{L\;c}(t)$ is constant when there is no production or decay of the asymmetry), we obtain in the absence of decays \cite{kls1} 
\be{e15} n_{L\;c}^{0}(t) = -\frac{\phi_{*}^{2} m_{\Phi} \sin(2 \theta)}{2}  \sin\left( \frac{2 A (t - t_{*}) }{m_{\Phi}} \right)  ~.\ee

In the limit where $\Gamma_{\Phi}^2 \ll m^2$ and $H^2 \ll m^2$, the inclusion of the decay terms multiplies the solutions for $\phi_{1}$ and $\phi_{2}$ by a factor $e^{-\Gamma_{\Phi} (t-t_{*})/2}$ and $n_{L\;c}^{0}$ by a factor $e^{-\Gamma_{\Phi} (t-t_{*})}$, therefore 
\be{e15a} n_{L\;c}(t) = - \frac{\phi_{*}^{2} \sin(2 \theta) m_{\Phi} e^{-\Gamma_{\Phi} (t - t_{*}) }}{2}   \sin\left( \frac{2 A (t - t_{*}) }{m_{\Phi}} \right)   ~.\ee
The lepton asymmetry transferred to the SM sector by inflaton decays is then given by  
\be{e15b} \hat{n}_{L\;c}(t) = \int_{t_{*}}^{t} \Gamma_{\Phi} n_{L\; c}(t)  dt  = - \frac{\Gamma_{\Phi} \phi_{*}^{2} \sin(2 \theta) m_{\Phi}}{2} \int_{t_{*}}^{t} e^{-\Gamma_{\Phi} (t - t_{*})} \sin\left( \frac{2 A (t - t_{*}) }{m_{\Phi}} \right) dt   ~, \ee
where `hat' denotes the transferred asymmetry. 
The total comoving asymmetry transferred to the SM sector in the limit $t \rightarrow \infty$
is then 
\be{e21}  \hat{n}_{L\;c,\;tot} = \frac{-\Gamma_{\Phi} \phi_{*}^{2} \sin(2 \theta) m_{\Phi}^{2}}{2 A} \left(1 + \left(\frac{\Gamma_{\Phi} m_{\Phi}}{2 A}\right)^{2} \right)^{-1}  ~.\ee

\begin{figure}[h]
\begin{center}
\hspace*{-0.5cm}\includegraphics[trim = -3cm 0cm 0cm 0cm, clip = true, width=0.55\textwidth, angle = -90]{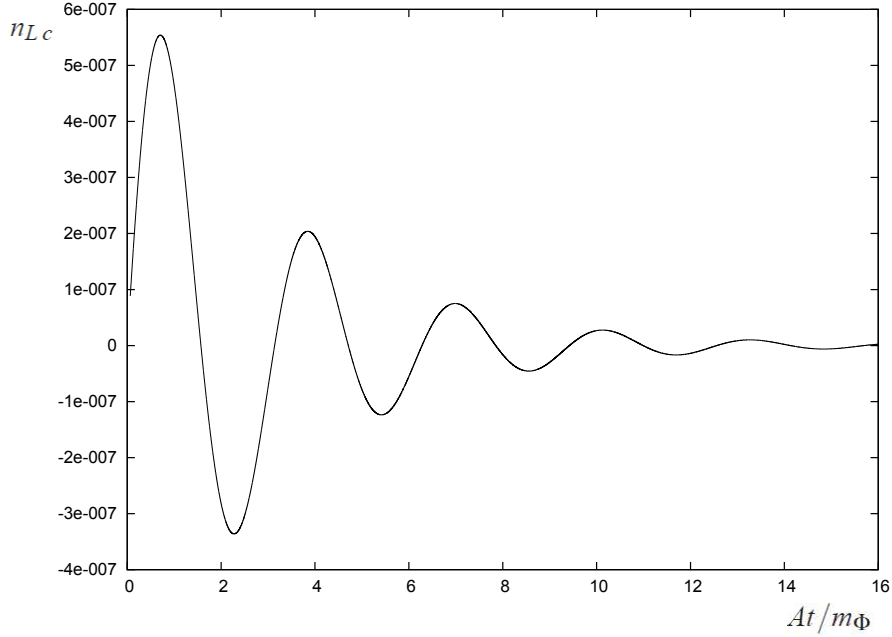}
\caption{Numerical comoving condensate lepton asymmetry for the case $\tau_{\Phi} = T_{asy}$, with $m_{\Phi} = 10^{16} \GeV$, $A^{1/2} = 10^{13} \GeV$, $\lambda_{\Phi} = 0.1$ and $\sin(2 \theta) = -1$,  illustrating the oscillations of the condensate asymmetry about zero.} 
\label{fig1}
\end{center}
\end{figure}

\begin{figure}[h]
\begin{center}
\hspace*{-0.5cm}\includegraphics[trim = -3cm 0cm 0cm 0cm, clip = true, width=0.55\textwidth, angle = -90]{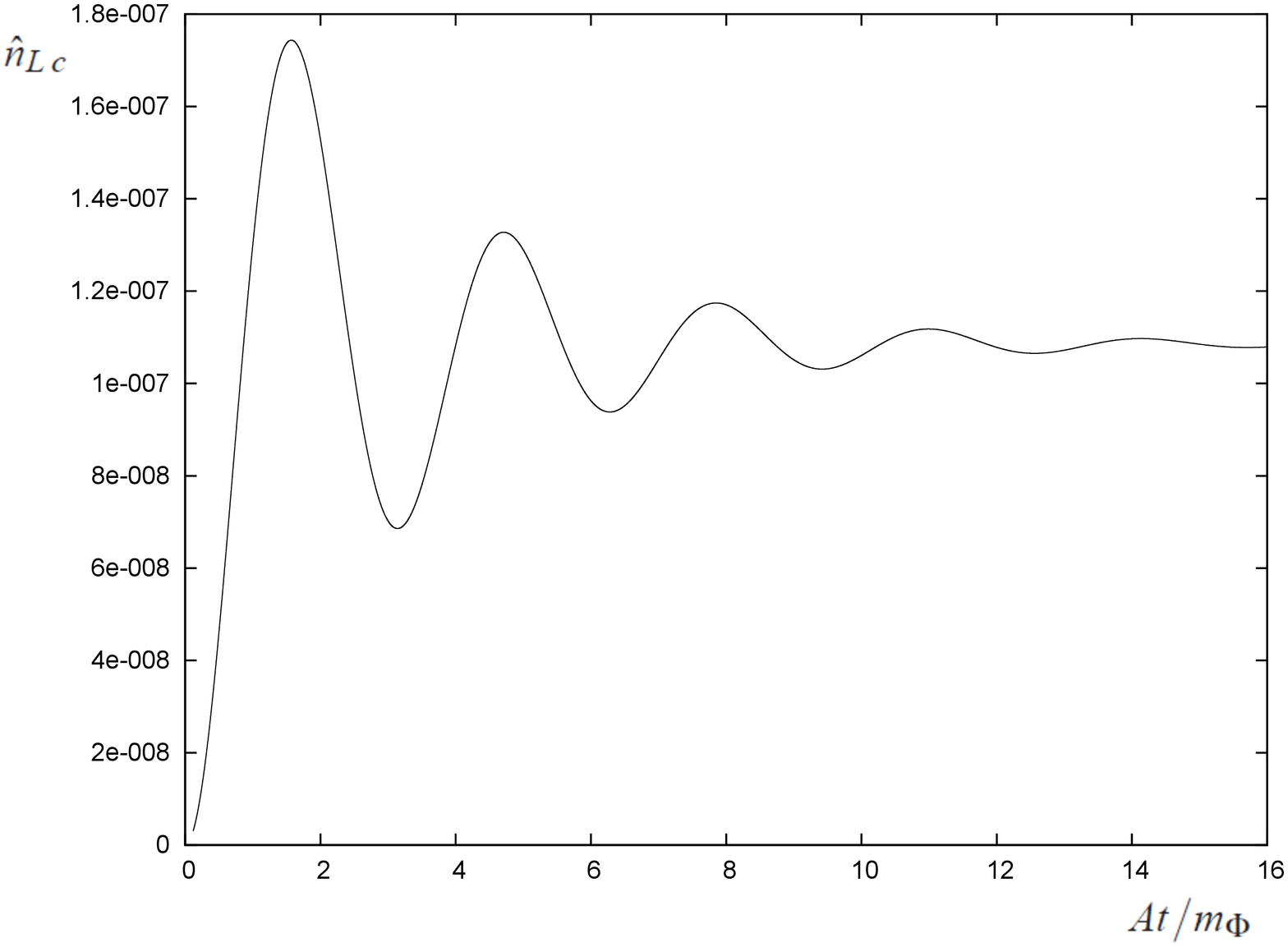}
\caption{Comoving transferred lepton asymmetry for the case $\tau_{\Phi} = T_{asy}$, showing the non-zero net transferred asymmetry.} 
\label{fig2}
\end{center}
\end{figure}

\begin{figure}[h]
\begin{center}
\hspace*{-0.5cm}\includegraphics[trim = -3cm 0cm 0cm 0cm, clip = true, width=0.55\textwidth, angle = -90]{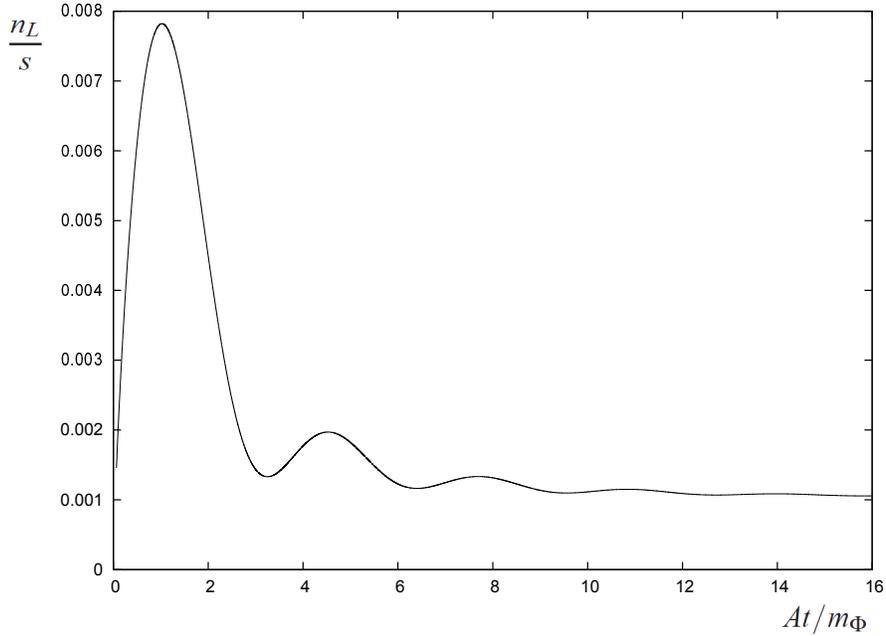}
\caption{The lepton-number to entropy ratio for the case $\tau_{\Phi} = T_{asy}$. The late-time final value almost coincides with the analytical prediction from \eq{e26b}. } 
\label{fig3}
\end{center}
\end{figure}

\begin{figure}[h]
\begin{center}
\hspace*{-0.5cm}\includegraphics[trim = -3cm 0cm 0cm 0cm, clip = true, width=0.55\textwidth, angle = -90]{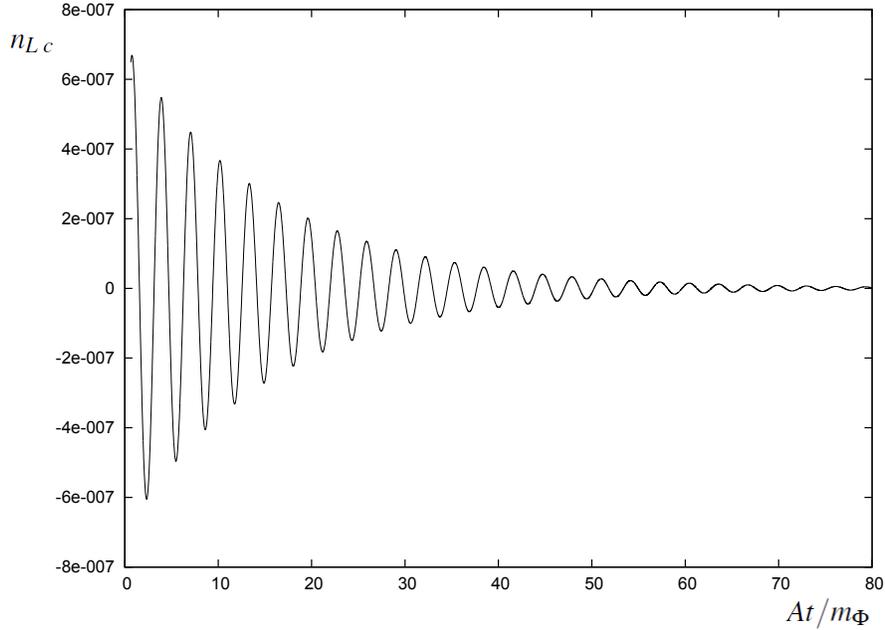}
\caption{Numerical comoving condensate lepton asymmetry for the case $\tau_{\Phi} = 5 T_{asy}$, with $m_{\Phi} = 10^{16} \GeV$, $A^{1/2} = 10^{13} \GeV$, $\lambda_{\Phi} = 0.1$ and $\sin(2 \theta) = -1$ and $T_{R} = 2.1 \times 10^{13} \GeV$, illustrating more rapid oscillations of the condensate asymmetry.} 
\label{fig4}
\end{center}
\end{figure}

\begin{figure}[h]
\begin{center}
\hspace*{-0.5cm}\includegraphics[trim = -3cm 0cm 0cm 0cm, clip = true, width=0.55\textwidth, angle = -90]{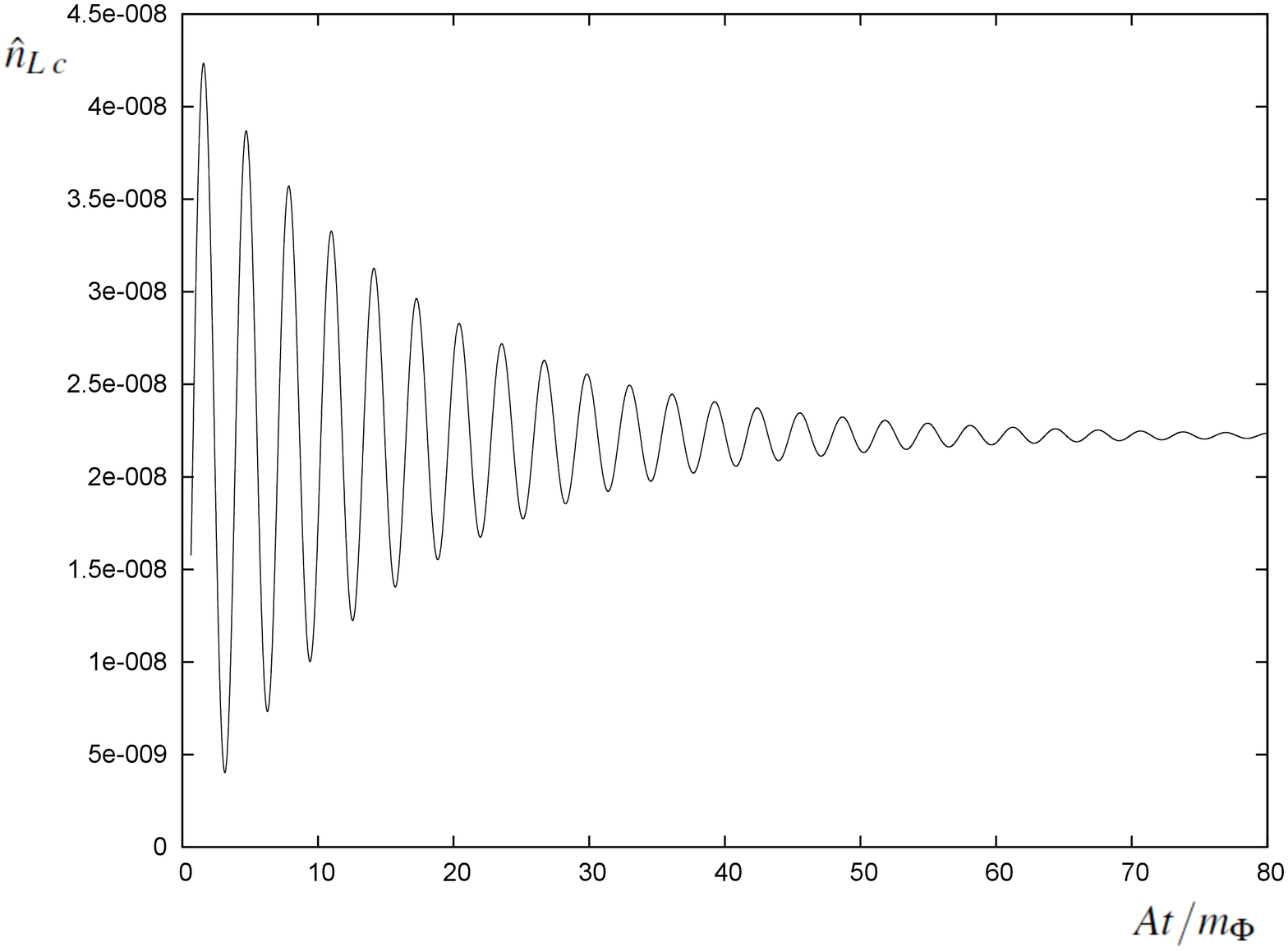}
\caption{Comoving transferred lepton asymmetry for the case $\tau_{\Phi} = 5 T_{asy}$, showing the non-zero net transferred asymmetry.} 
\label{fig5}
\end{center}
\end{figure}

\begin{figure}[h]
\begin{center}
\hspace*{-0.5cm}\includegraphics[trim = -3cm 0cm 0cm 0cm, clip = true, width=0.55\textwidth, angle = -90]{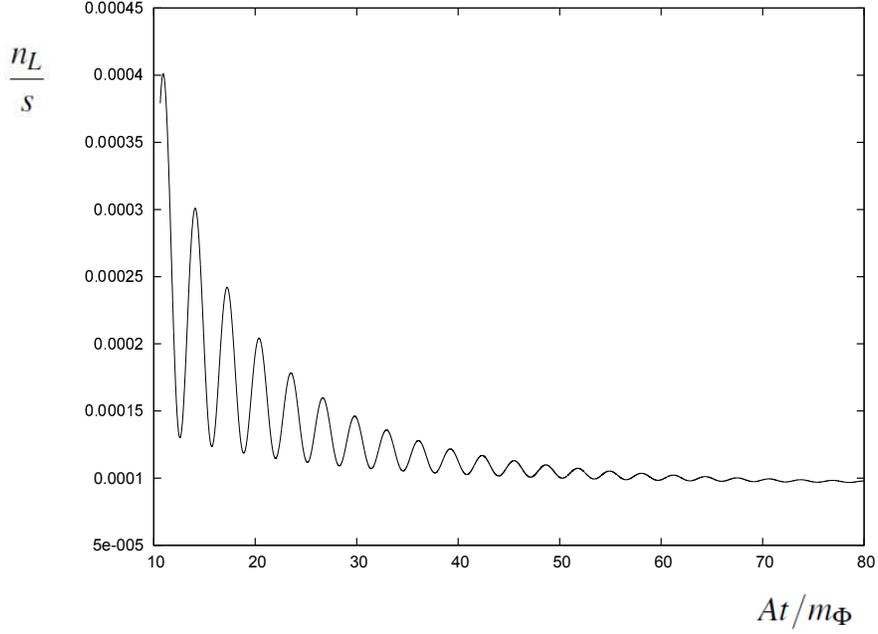}
\caption{The lepton-number to entropy ratio for the case $\tau_{\Phi} = 5 T_{asy}$. The late-time final value almost coincides with the analytical prediction from \eq{e26b}. (We have shown the values of $n_{L}/s$ starting from $At/m_{\Phi} = 10$ in order to emphasize the late-time value.)  } 
\label{fig6}
\end{center}
\end{figure}

There are two regimes for $A$, corresponding to the cases where the inflaton lifetime $\tau_{\Phi}$ is short ($\Gamma_{\Phi} \gg 2A/m_{\Phi}$) and where it long ($\Gamma_{\Phi} \ll 2A/m_{\Phi}$) compared to $2A/m_{\Phi} \equiv T_{asy}/2 \pi$. The resulting transferred asymmetry is 
\be{n1}  \hat{n}_{c,\;tot} = \frac{-2 A \phi_{*}^{2} \sin(2 \theta)}{\Gamma_{\Phi}} \;\;\;\; \left(\tau_{\Phi} \ll \frac{T_{asy}}{2 \pi} \right)  ~\ee
and 
\be{n2}  \hat{n}_{c,\;tot} = \frac{- \Gamma_{\Phi} \phi_{*}^{2} m_{\Phi}^{2} \sin(2 \theta)}{2 A}   \;\;\;\; \left(\tau_{\Phi} \gg \frac{T_{asy}}{2 \pi} \right)  ~.\ee
The maximum possible lepton asymmetry is at $\Gamma_{\Phi} m_{\Phi}/2A = 1$, corresponding to $\tau_{\Phi} = T_{asy}/2 \pi$,
\be{n3} \hat{n}_{c,\;max} = - \frac{m_{\Phi} \phi_{*}^{2} \sin(2 \theta) }{2}  ~.\ee

In order to have an analytical relation for the baryon number to entropy of the condensate,  we will use the approximation that the Universe is matter dominated until $\Gamma_{\phi} = H$, at which time the energy in the condensate and the total final lepton number asymmetry is instantly transferred to radiation and the SM sector lepton number asymmetry. We will check the accuracy of this approximation later in this section. In this case, the decay rate of the field is related to the reheating temperature $T_{R}$ by $\Gamma_{\Phi} = k_{T_{R}} T_{R}^{2}/M_{Pl}$, where $k_{T_{R}} = (\pi^{2} g(T_{R})/90)^{1/2}$ and $g(T)$ is the number of relativistic degrees of freedom. To convert total the comoving lepton asymmetry to the physical asymmetry at $T \leq T_{R}$ we multiply $\hat{n}_{L\; tot}$ by $(a_{*}/a(T_{R}))^{3} = H(T_{R})^{2}/H_{*}^{2} = 6 k_{T_{R}}^{2} T_{R}^{4}/m_{\Phi}^{2} \phi_{*}^{2}$. The final lepton number to entropy after inflaton decay at $T_{R}$, where the entropy density is given by $s = 4 k_{T}^{2} T^{3}$, is then
\be{e26b} \frac{n_{L}}{s} = -\frac{3}{4} \frac{k_{T_{R}} T_{R}^{3}\sin \left(2 \theta\right)  }{A M_{Pl} } \;\;\;\; \left (\tau_{\Phi} \gg \frac{T_{asy}}{2 \pi} \right)                                        ~\ee
and
\be{n4} \frac{n_{L}}{s} = -\frac{3 A M_{Pl}\sin(2 \theta)}{k_{T_{R}} T_{R} m_{\phi}^{2}} \;\;\;\; \left(\tau_{\Phi} \ll \frac{T_{asy}}{2 \pi} \right)                                     ~.\ee 
The maximum possible asymmetry is 
\be{n5} \frac{n_{L,\; max}}{s} = -\frac{3 T_{R} \sin \left(2 \theta\right)  }{4 m_{\Phi}} ~.\ee

The transferred lepton asymmetry is partially converted to a baryon asymmetry via $(B + L)$-violating 
sphaleron fluctuations, with the final baryon asymmetry to entropy ratio $n_{B}/s$ given by \cite{sph,sph2}
\be{n6}  \frac{n_{B}}{s} = - \frac{28}{79}\frac{\hat{n}_{L}}{s}   ~.\ee 
The final baryon asymmetry is then 
\be{n7} \frac{n_{B}}{s}   = 3.7 \times 10^{-21}   \,\frac{m_{\Phi}^{2}}{A}   \left( \frac{T_{R}}{10^{8} \GeV} \right)^{3}  \left( \frac{10^{13} \GeV}{m_{\Phi}} \right)^{2}  \sin \left(2 \theta\right)  \;\;\;\; \left(\tau_{\Phi} \gg \frac{T_{asy}}{2 \pi} \right)                                                                            ~\ee
and
\be{n8} \frac{n_{B}}{s} =  7.7 \times 10^{9} \, \frac{A}{m_{\Phi}^{2}} \left( \frac{10^{8} \GeV}{T_{R}}\right) \sin \left(2 \theta\right)  \;\;\;\; \left(\tau_{\Phi} \ll \frac{T_{asy}}{2 \pi} \right)                                                                            ~,\ee
where  we have normalised the expression to some representative values of $T_{R}$ and $m_{\Phi}$.The maximum possible asymmetry is  
\be{n9}  \frac{n_{B,\; max}}{s}  = 2.7 \times 10^{-6} \, \left( \frac{T_{R}}{10^{8} \GeV} \right)  \left( \frac{10^{13} \GeV}{m_{\Phi}} \right) \sin \left(2 \theta\right)    ~.\ee

The observed baryon-to-entropy ratio is  
$ (n_{B}/s)_{obs}  = 0.861 \pm 0.005 \times 10^{-10} $.  
In order to account for the observed asymmetry, we therefore require that
\be{e40} \frac{A^{1/2}}{m_{\Phi}} = 6.7 \times 10^{-6} \sin^{1/2} \left(2 \theta\right)    
\left( \frac{10^{13} \GeV }{m_{\Phi}} \right) 
\left( \frac{T_{R}}{10^{8} \GeV} \right)^{3/2} \;\;\;\; \left(\tau_{\Phi} \gg \frac{T_{asy}}{2 \pi} \right)                                                                            
~\ee
and
\be{ea2} \frac{A^{1/2}}{m_{\Phi}} = 1.1 \times 10^{-10} \,\left( \frac{T_{R}}{10^{8} \GeV} \right)^{1/2}
\left(\frac{1}{\sin \left(2 \theta\right)} \right)^{1/2}   \;\;\;\; \left(\tau_{\Phi} \ll \frac{T_{asy}}{2 \pi} \right)                                                                            
~.\ee

\subsection{Numerical Lepton Asymmetry}

 In order to check our analytical expressions, we have solved the complete field equations and included the continuous transfer of energy from the decaying inflaton condensate to radiation  via 
\be{nx1}  \frac{d \rho_{rad}}{dt} +  4 H \rho_{rad} = \Gamma_{\Phi} \rho_{\Phi}   ~.\ee 
In the figures we show the exact numerical solution for two cases, $\tau_{\Phi} = T_{asy}$ and $\tau_{\Phi} = 5 T_{asy}$. In Figures 1-3 we show the evolution of the comoving condensate asymmetry, the comoving transferred asymmetry, and the lepton number-to-entropy for the case $\tau_{\Phi} = T_{asy}$, with $m_{\Phi} = 10^{16} \GeV$, $A^{1/2} = 10^{13} \GeV$ and  $\lambda_{\Phi} = 0.1$. (The comoving asymmetries are given in Planck units.) For this case  $T_{R} = 4.8 \times 10^{13} \GeV$ using the instantaneous decay expression. In Figures 4-6 we show the corresponding results for the case with $\tau_{\Phi} = 5 T_{asy}$, in which case $T_{R} = 2.1 \times 10^{13} \GeV$. In order to test the analytical predictions we have chosen values of $m_{\Phi}$ and $\Gamma_{\Phi}$ which allow for numerical solution of the field equations without extreme differences between the field oscillation time and the $\Phi$ lifetime, even though these values do not produce realistic lepton asymmetries.  

The analytical prediction for the lepton number-to-entropy ratio  from \eq{e26b} for the case $\tau_{\Phi} = T_{asy}$ is 
\be{n4a} \left(\frac{n_{L}}{s}\right)_{analytical} =  1.15 \times 10^{-3}    ~, \ee
where we have used $g(T_{R}) = 100$ and set $\sin(2 \theta) = -1$ to have a positive lepton asymmetry. The corresponding numerical result from Figure 3 is
\be{n4b}  \left(\frac{n_{L}}{s}\right)_{numerical}  =  1.106 \times 10^{-3}    ~.\ee
The analytical prediction for the lepton number-to-entropy ratio  for the case $\tau_{\Phi} = 5 T_{asy}$ is 
\be{n4a} \left(\frac{n_{L}}{s}\right)_{analytical}  =  1.03 \times 10^{-4}    ~.\ee
The corresponding numerical result from Figure 6 is
\be{n4b}  \left(\frac{n_{L}}{s}\right)_{numerical}  =  9.80 \times 10^{-5}    ~.\ee
Therefore the analytical solution gives a very accurate estimate of the final lepton-to-entropy ratio.   

\subsection{Consistency of the assumption that AD leptogenesis occurs during $|\Phi|^2$ domination} 

   The above analysis assumes that the lepton asymmetry is entirely generated when the potential is dominated by the $|\Phi|^2$ term.  The condition for this assumption to be valid is 
that the $A$-terms do not significantly influence the motion of the $\phi_{1}$ and $\phi_{2}$ fields until $\phi < \phi_{*}$, when the potential is dominated by the $|\Phi|^{2}$ term. This is true if the mass of the angular field perturbations about the minimum of the potential as a function of $\theta$ for a given $\phi$ is less than $H$ when $\phi \geq \phi_{*}$. This can be understood by considering the field equations in terms of the radial and angular variables $\Phi = \phi e^{i \theta}/\sqrt{2}$. 
\be{cd1} \ddot{\phi} + 3 H \dot{\phi} = -(m_{\Phi}^{2} + 2 \lambda_{\Phi} |\Phi|^{2} ) \phi - 2 A \phi \cos (2 \theta)   ~\ee 
and
\be{cd2} \phi \ddot{\theta} + 2 \dot{\theta} \dot{\phi} + 3 H \phi \dot{\theta} = 2 A \phi \sin (2 \theta)   ~.\ee 
For $A$ real and positive, the minimum of the potential is along the direction $\theta = \pi/2$. To estimate the condition for the angle $\theta$ to be unaffected during the $\phi$ oscillations, we can consider a small perturbation of $\theta$ from the minimum direction. For convenience, we change variable to $\hat{\theta} =  \pi/2 - \theta$ so that the minimum direction becomes $\hat{\theta} = 0$. The equations then become 
\be{cd1a} \ddot{\phi} + 3 H \dot{\phi} = -(m_{\Phi}^{2} + 2 \lambda_{\Phi} |\Phi|^{2} ) \phi + 2 A \phi \cos (2 \hat{\theta})   ~\ee 
and
\be{cd2a} \phi \ddot{\hat{\theta}} + 2 \dot{\hat{\theta}} \dot{\phi} + 3 H \phi \dot{\hat{\theta}} = - 2 A \phi \sin (2 \hat{\theta})   ~.\ee 
Then, for $\hat{\theta} \ll 1$, we have 
\be{cd3} \ddot{\phi} + 3 H \dot{\phi} = -(m_{\Phi}^{2} - 2 \lambda_{\Phi} |\Phi|^{2} - 2A ) \phi  ~\ee 
and
\be{cd4} \phi \ddot{\hat{\theta}} + \left(3 H + 2 \frac{\dot{\phi}}{\phi} \right)  \phi \dot{\hat{\theta}} = 4 A \phi \hat{\theta}   ~.\ee  
From the RHS of \eq{cd4}, if we average over the $\phi$ oscillations and consider $\phi$ to be a constant, and define a field $\sigma = \phi \hat{\theta}$, then the mass of the $\sigma$ field is $m_{\sigma} = \sqrt{4 A} $. 

So we can expect the dynamics of the angular field to be unaffected by the $A$-term if $m_{\sigma}^{2} < H^{2}$. If true,  we can consider effectively $A = 0$ in \eq{cd4} and so $\theta$ equal to a constant will be a solution. Therefore, as long as $4A < H^2$, the AD dynamics due to the $A$-term will be negligible. We therefore require that this is satisfied at $\phi \gae \phi_{*} = m_{\Phi}/\sqrt{\lambda_{\Phi}}$. Let $\phi_{AD}$ be the value of the field at which $4 A = H^{2}$ and the phase field evolution (and so Affleck-Dine dynamics) becomes important. We then require that $\phi_{AD} < \phi_{*}$, which requires that 
\be{cd5} 4 A < H_{*}^{2}   ~.\ee 
At $\phi = \phi_{*}$, $V(\phi_{*}) = m_{\Phi}^{2} \phi_{*}^{2}/2 + \lambda_{\Phi} \phi_{*}^{4}/4 =3 m_{\Phi}^{4}/4 \lambda_{\Phi}$ and $H_{*} = (V(\phi_{*})/3 M_{Pl}^{2})^{1/2} = m_{\Phi}^{2}/2 \sqrt{\lambda_{\Phi}} M_{Pl}$. The condition \eq{cd5} then becomes 
\be{cd6}  \frac{A^{1/2}}{m_{\Phi}} < \frac{A_{th}^{1/2}}{m_{\Phi}} = \frac{m}{4 \sqrt{\lambda_{\Phi}} M_{Pl}} = 1.04 \times 10^{-6} \lambda_{\Phi}^{-1/2} \left(\frac{m_{\Phi}}{10^{13} \GeV}\right)   ~,\ee
where $A_{th}$ is the value of $A$ below which the threshold approximation is valid.     
For $A < A_{th}$, the asymmetry will be entirely generated during $\phi^{2}$ oscillations and the analytical approximation is valid. In \cite{kls1} it was confirmed numerically that if \eq{cd6} is satisfied then the analytical prediction for the asymmetry based on $|\Phi|^{2}$ domination is valid. If this condition is not satisfied and the $L$-violating mass terms become dynamical during the $\phi^4$ oscillation-dominated era, then there is an additional suppression of the asymmetry due to the evolution of the phase during $\phi^{4}$-dominated oscillations \cite{kls1}.

\subsection{Consistency with Non-Minimally Coupled Inflation}

In our calculation of the asymmetry we have assumed that the effect of the non-minimal coupling is negligible. This requires that the $\phi^{2}$-dominated evolution begins after the modification of the dynamics due to the non-minimal coupling becomes negligible. We therefore require that $\phi_{*} < \phi_{c}$, where $\phi_{c}$ is the value of the inflaton field at which the non-minimal coupling strongly modifies the dynamics. For the metric case $\phi_{c}$ is given by \eq{i18} and for the Palatini case by \eq{i19}.

The condition $\phi_{*} = m_{\Phi}/\sqrt{\lambda_{\Phi}} < \phi_{c}$ then gives for the metric case 
\be{cd9} m_{\Phi} < \frac{\sqrt{\lambda_{\Phi}} M_{Pl}}{\sqrt{6} \xi}  ~\ee 
and for the Palatini case
\be{cd10} m_{\Phi}  < \frac{\sqrt{\lambda_{\Phi}} M_{Pl}}{\sqrt{\xi}}   ~.\ee
Using the expressions \eq{i16} and \eq{i17} for $\xi$ at $N = 55$, which give $\xi = 4.5 \times 10^{4} \sqrt{\lambda_{\Phi}}$ for the metric case and $\xi = 1.2 \times 10^{10} \,\lambda_{\Phi}$ for the Palatini case, we obtain the same condition for both the metric and Palatini cases  
\be{cd11}  m_{\Phi} < 2.2 \times 10^{13} \GeV  ~.\ee

\section{Lepton Asymmetry Transfer, Reheating and Washout}

\subsection{Lepton Asymmetry Transfer and Reheating}

We assume that reheating occurs through the perturbative decay of the condensate scalars to RH neutrinos.   In this case, reheating and lepton number transfer to the SM sector occur simultaneously. The inflaton decay rate for $\Phi \rightarrow N_{R}N_{R}$ is 
\be{r1} \Gamma_{\Phi} = \frac{y_{\Phi}^{2} m_{\Phi}}{8 \pi}    ~.\ee 
As usual, we define $T_{R}$ to be the effective temperature at which the Universe becomes dominated by relativistic particles after $\Phi$ decay, in this case right-handed neutrinos. The RH neutrinos will subsequently decay to light SM particles via $N_{R} \rightarrow HL$, at which time the Universe rapidly thermalises to a temperature $T_{th} \leq T_{R}$.  If the $N_{R}$ decay rate is fast enough that they decay immediately after being produced, then $T_{th} = T_{R}$. 

 As a specific example to show that reheating and lepton asymmetry transfer can easily be achieved without washout occurring due to L-violating scattering processes, we consider the case with $m_{\Phi} = 10^{13} \GeV$ and $T_{R} = 10^{8} \GeV$, with $m_{N}$ assumed to be small compared to the inflaton mass but large compared to the reheating temperature, $m_{N} = 10^{10} \GeV$. In this case the number density of RH neutrinos in thermal equilibrium is strongly Boltzmann suppressed. The reheating temperature is determined by $\Gamma_{\Phi} = H(T_{R}) =  k_{T_{R}} T_{R}^{2}/M_{Pl}$.
The coupling $y_{\Phi}$ required for a given effective decay temperature is then
\be{r3}  y_{\Phi} = 1.9 \times 10^{-7} \left( \frac{10^{13} \GeV}{m_{\Phi}} \right)^{1/2} 
\left( \frac{T_{R}}{10^{8} \GeV} \right)    ~.\ee  
The rest frame $N_{R} \rightarrow HL$ decay rate is  
\be{r4} \Gamma_{N_{R}} = \frac{h_{\nu}^{2} m_{N}}{8 \pi} 
= \frac{m_{\nu} m_{N}^{2}}{4 \pi v^{2} }    ~,\ee
where we have used \eq{e4}.
The energy of $N_{R}$ immediately after pair production will be $E_{N} = m_{\Phi}/2$. Assuming that $m_{\Phi} \gg m_{N}$, the produced $N_{R}$ are relativistic, $E_{N} \gg m_{N}$, and so the decay rate of the $N_{R}$ will be reduced by the time-dilation factor $m_{N}/E_{N}$. The condition that $N_{R}$ can decay immediately after production is then 
\be{r5}   \frac{m_{N}}{E_{N}} \Gamma_{N_{R}} > H(T_{R})  \Rightarrow \frac{m_{\nu} m_{N}^{3}}{2 \pi v^{2} m_{\Phi} } > \frac{k_{T} T_{R}^{2}}{M_{Pl}} ~.\ee
This is satisfied if 
\be{r6}   m_{\nu} > 5.2 \times 10^{-5} \eV \times 
\left( \frac{m_{\Phi}}{10^{13} \GeV} \right)  
 \left( \frac{10^{10} \GeV}{m_{N}} \right)^{3} 
 \left( \frac{T_{R}}{10^{8} \GeV} \right)^{2}  ~.\ee 
Thus if we consider $m_{\nu}$ to be of the order of the observed neutrino mass splittings, $m_{\nu} \sim 0.01 - 0.1 \eV$, then in this example both reheating and the thermalisation of the SM thermal background will occur immediately when the inflaton decays, since the RH neutrinos immediately decay to SM particles. 

\subsection{Absence of Washout} 

We next consider the condition for the SM lepton asymmetry  to avoid washout after being transferred from the inflaton condensate. Processes involving thermal background RH neutrinos will be very strongly Boltzmann suppressed, since $m_{N}/T_{R} = 100$ for this example. $\Delta L = 2$ scattering processes can occur via heavy $N$ exchange between $LH$. The resulting Weinberg operator is 
\be{r7}   \frac{h_{\nu}^{2}}{M_{N}}LHLH   ~\ee 
and the thermal scattering rate is \cite{wein}  
\be{r8} \Gamma_{W} = \frac{h_{\nu}^{4}}{4 \pi} \frac{T^{3}}{m_{N}^{2}} = \frac{m_{\nu}^{2} T^{3}}{\pi v^{4}}    ~,\ee
where have used the relation \eq{e4}.   
Requiring that this is less than $H$ at reheating gives the bound 
\be{r10} T_{R} < \frac{k_{T} \pi v^{4}}{m_{\nu}^{2} M_{Pl}} = 
1.6 \times 10^{12} \left(\frac{0.1 \eV}{m_{\nu}} \right)^{2} \GeV ~.\ee
Therefore, for $m_{N} = 10^{10} \GeV$, $T_{R} = 10^{8} \GeV$ and $m_{\nu} \lae 0.1 \eV$, there is no danger of washout of the lepton asymmetry after it transfers to the SM sector.   

Thus with $m_{\Phi} = 10^{13}$ GeV, $m_{N} = 10^{10} \GeV$ and $T_{R} = 10^{8}$ GeV, the inflaton condensate will decay to relativistic RH neutrinos pairs which immediately decay to $LH$ pairs and thermalise the SM thermal background.  Thermal L-violating scattering processes are ineffective, therefore the lepton asymmetry successfully transfers to the SM sector without danger of washout. The final baryon asymmetry is produced by $(B + L)$-violating sphaleron fluctuations, which partially convert the lepton asymmetry to a baryon asymmetry. 

We note that reheating temperature in this example, $T_{R} = 10^{8} \GeV$, is within the range $10^{6}-10^{8} \GeV$ that may be observable in the spectrum of primordial gravitational waves with $r = 0.001 - 0.1$ \cite{gravrh}. This is in contrast to the case of thermal leptogenesis, which requires that $T_{R} \gae 10^{9} \GeV$ \cite{lept,lept2,lept3}. For the metric non-minimally coupled inflation model, the tensor-to-scalar index is $r = 0.004$, which is large enough to be observed by next generation telescopes such as LiteBird \cite{ltb}. Therefore the model could explain both the observed baryon asymmetry and a future observation of primordial gravitational waves with a reheating signature in their spectrum.

\subsection{Condensate Decay via the Higgs portal, via the Non-Minimal Coupling and via Gravitational Decay}

\subsubsection{Decay via the Higgs Portal} 

In the above analysis, we have considered reheating to occur through perturbative decays of the inflaton to right-handed neutrinos. In general, we expect a Higgs portal coupling of the inflaton to the Higgs doublet bosons to exist, as it cannot be suppressed by any symmetry. This can allow condensate decay via annihilation of the condensate scalars. Here we will estimate an upper limit on the coupling of the inflaton to the Higgs boson doublet for condensate decay to right-handed neutrinos to dominate reheating. 

The portal coupling is  
\be{hh1}  \lambda_{\Phi H} |\Phi|^{2} |H|^{2} =  
\frac{\lambda_{\Phi H}}{4}  \phi^{2} \sum_{i = 1}^{4} h_{i}^{2}  ~,\ee
where $h_{i}$ are the real scalars of the Higgs doublet. 
To estimate the decay of the condensate to the Higgs doublet scalars, we will use the results given in Appendix A of \cite{cond}. For a single real Higgs scalar in \eq{hh1}, the condensate decay rate has the form 
\be{hh2} \Gamma_{portal} = \frac{C\,\lambda_{\Phi H}^{2} \rho_{\phi}}{256 \pi m_{eff}^{3}}  ~,\ee 
where the effective mass of the condensate scalars is $m_{eff}^{2} = V^{''}(\phi)$ and $\rho_{\phi} \equiv V(\phi)$, where $\phi$ is the amplitude of oscillation. The constant $C$ depends on the potential. For the $\phi^2$ potential, $m_{eff} = m_{\phi}$  and $C = 1$. For the $\phi^4$ potential, $m_{eff} = \sqrt{3 \lambda_{\Phi}} \phi$ and $C \approx 18$ \cite{cond}.

We wish to ensure that the condensate decay rate via the Higgs portal is less than $H$ for all $\phi$. We first derive the condition on $\lambda_{\Phi H}$ for this to be true at $\phi < \phi_{*}$.  We then show that if decay does not occur at $\phi < \phi_{*}$ then it will not occur at $\phi > \phi_{*}$. 

At $\phi < \phi_{*}$ we have a $\phi^{2}$ potential. Therefore $\rho_{\phi} = m_{\Phi}^{2} \phi^{2}/2$ and 
\be{hh3} \Gamma_{portal} \approx \frac{\lambda_{\Phi H}^{2} \phi^{2}}{128 \pi m_{\Phi}}  ~,\ee
where we have summed over all four real scalars of the Higgs doublet.  
The condition that $\Gamma_{portal} \lae H = m \phi/\sqrt{6} M_{Pl}$ at $\phi < \phi_{*}$ is strongest at $\phi = \phi_{*}$, where it requires that 
\be{hh4} \lambda_{\Phi H} \lae \left(\frac{128 \pi}{\sqrt{6}}\right)^{1/2} \left( \frac{m_{\Phi}}{M_{Pl}}\right)^{1/2} \lambda_{\Phi}^{1/4}   ~.\ee
Therefore
\be{hh5} 
\lambda_{\Phi H} \lae 2.6 \times 10^{-2} \, \lambda_{\Phi}^{1/4} \left( \frac{m_{\Phi}}{10^{13} \GeV}\right)^{1/2}  ~.\ee
Thus only a moderate suppression of the Higgs portal coupling is required in the example we are considering. 

At $\phi > \phi_{*}$  we have $m_{eff} \propto \phi$ and $V \propto \phi^{4}$. Therefore $\Gamma_{portal} \propto V(\phi)/m_{eff}^{3} \propto \phi$. Since $H \propto V^{1/2} \propto \phi^{2}$, it follows that if $\Gamma_{portal} < H$ at $\phi = \phi_{*}$, then this will also be true at $\phi > \phi_{*}$. Thus \eq{hh5} is sufficient to ensure that decay via the Higgs portal is negligible. 

If the portal coupling were to dominate reheating, then the symmetric component of the inflaton condensate could decay earlier, but the asymmetric component would remain in the form of a maximally asymmetric $\Phi$ condensate, with a circular orbit in the field space, until decay to RH neutrinos occurred. As a result, there could be two separate reheating events, with initial reheating leaving a maximally asymmetric inflaton condensate, which could later come to dominate the radiation density and eventually decay to RH neutrinos, reheating the Universe a second time. 

\subsubsection{Decay via the Non-Minimal Coupling}  

The inflaton has a large non-minimal quadratic coupling to $R$, suggesting that annihilations to SM particles via the non-minimal coupling could be important. To estimate the decay rate of the condensate to SM particles, we will extend the non-minimally coupled action to include the SM Higgs doublet. 

In the Einstein frame, even in the absence of perturbative interactions between the inflaton and SM sectors, the non-minimal coupling will create non-renormalisable interactions. The important terms in the Einstein frame Lagrangian are 
\be{c1} -\frac{1}{\Omega^{2}} \partial_{\mu} H^{\dagger} \partial^{\mu} H - \frac{\lambda_{h} |H|^{4}}{\Omega^4}  ~.\ee 
During $\phi$ oscillations after inflation we have $\xi \phi^{2}/M_{Pl}^{2} \ll 1$ and we can therefore expand the conformal factors. We then obtain for the leading order non-renormalisable interactions between $\phi$ and each real Higgs scalar $h_{i}$, 
\be{c2}  \frac{\xi}{2 M_{Pl}^{2}} \phi^{2} \sum_{i=1}^{4}\partial_{\mu}h_{i} \partial^{\mu} h_{i} + \frac{\lambda_{h} \xi}{2 M_{Pl}^{2}} \phi^{2} \left(\sum_{i = 1}^{4} h_{i}^{2}\right)^{2}   ~.\ee 
These interactions allow the scalars in the condensate to annihilate to Higgs scalars.

To estimate the annihilation rate, we will consider the first interaction in \eq{c2}; the second should give a similar annihilation rate, having a suppression due to $\lambda_{h}$ but an enhancement due to the larger phase space for decay to 4$h_{i}$. Since we have annihilation of zero momentum scalars, the final state Higgs will have energy $E_{h} = m_{\Phi}/2$ and  momenta $k_{1}$ and $k_{2}$ such that $k_{1}.k_{2} = 2 E_{h}^{2} = m_{\Phi}^{2}/2$. We can therefore replace the first term in \eq{c2} by an effective portal coupling 
\be{c3} \sum_{i = 1}^{4} \frac{\xi m_{\Phi}^{2}}{4 M_{Pl}^{2}} \phi^{2} h_{i}^{2}   ~.\ee 
We can then apply the previous analysis of the portal coupling 
with $\lambda_{\Phi H} \rightarrow \xi m_{\Phi}^{2}/M_{Pl}^{2}$. 
From \eq{hh4} the condition for annihilations to Higgs bosons via the non-minimal coupling to be negligible is 
\be{c4} m_{\Phi} \lae \frac{\lambda_{\Phi}^{1/6}}{\xi^{2/3}} \left(\frac{128 \pi}{\sqrt{6}}\right)^{1/3}  M_{Pl}   ~\ee 
For metric inflation with $N = 55$ we have $\xi = 4.5 \times 10^{4} \sqrt{\lambda_{\Phi}}$. Therefore \eq{c4} becomes 
\be{c5} m_{\Phi} \lae \frac{1.0 \times 10^{16} \GeV}{\lambda_{\Phi}^{1/6}}   ~.\ee 
For Palatini inflation with $N = 55$ we have $\xi = 1.2 \times 10^{10} \lambda_{\Phi}$ and \eq{c4} becomes 
\be{c6} m_{\Phi} \lae \frac{2.5 \times 10^{12} \GeV}{\lambda_{\Phi}^{1/2}}   ~.\ee 

The bound on $m_{\Phi}$ is easily satisfied for metric inflation when $m_{\Phi}$ is small enough for non-minimal dynamics to be negligible, $m_{\Phi} < 2.2 \times 10^{13} \GeV$. For Palatini inflation, a mild suppression of $\lambda_{\Phi}$, with $\lambda_{\Phi}$ less than about 0.01, would be necessary to satisfy \eq{c6} at the dynamical upper limit of $m_{\Phi}$.

\subsubsection{Gravitational Decay of the Condensate} 

Assuming that the $\phi \leftrightarrow -\phi$ symmetry of the potential is unbroken by gravitational interactions, any purely gravitational decay of the condensate will involve graviton exchange processes with $\phi^{2}$ and therefore annihilation of the condensate scalars to SM particles. In this case we expect that annihilation via the non-minimal coupling, which is characterised by a mass scale $M_{Pl}/\sqrt{\xi}$, will easily dominate any purely gravitational annihilation mode scaled by $M_{Pl}$. Therefore we expect that direct gravitational decay of the condensate will generally be negligible if decay via the non-minimal coupling is negligible. 
 
It is known that gravitational instantons can break global symmetries, which could lead to $U(1)_{L}$- and $Z_{2}$-breaking couplings of the inflaton to SM particles. However, being a non-perturbative effect, the strength of the couplings is strongly dependent upon the UV completion of gravity, with an exponential suppression $e^{-S}$ due to the instanton tunnelling action $S$ in addition to suppression by the Planck scale. Therefore there is no reason to assume that this effect will significantly modify reheating.

\section{Baryon Isocurvature Perturbations} 

We next consider the possibility of observable baryon isocurvature perturbations in this model. For $A^{1/2} < m_{\Phi} \ll H$ during inflation, the phase $\theta$ of $\Phi$ is effectively a massless field. Therefore quantum fluctuations of $\theta$ will give rise to baryon isocurvature perturbations. The magnitude of the isocurvature perturbations will depend upon the specific non-minimally coupled inflation model being considered: metric or Palatini. In non-minimally coupled inflation models in the Einstein frame, the kinetic term for $\phi_{2}$ (where $\Phi = (\phi_{1} + i \phi_{2})/\sqrt{2}$) is  
\be{r12}   \frac{1}{2 \Omega^{2}} \partial_{\mu} \phi_{2} \partial^{\mu} \phi_{2}    ~,\ee   
where 
\be{r13} \Omega^{2} = 1 + \frac{\xi (\phi_{1}^{2} + \phi_{2}^{2})}{M_{Pl}^{2}}   ~.\ee 
During inflation, the inflaton, which we choose to be  $\phi_{1}$, has a slow-rolling background value  $\overline{\phi}_{1}$, whilst by defining $\theta$ to be zero along $\phi_{1}$ we can consider $\phi_{2}$ to be purely due to quantum fluctuations. Treating $\overline{\phi}_{1}$ as effectively constant on the time scales over which the $\phi_{2}$ quantum fluctuations are produced, and with $\xi \overline{\phi}_{1}^{2}/M_{Pl}^{2} \gg 1$ during inflation, the $\phi_{2}$ kinetic term is 
\be{r14} \frac{M_{Pl}^{2}}{2 \xi \overline{\phi}_{1}^{2}}  \partial_{\mu} \phi_{2} \partial^{\mu} \phi_{2}   ~.\ee
The fluctuations of the phase $\theta$ are related to the fluctuations of $\phi_{2}$ by 
\be{r14a} \Phi = \frac{\phi}{\sqrt{2}}e^{i \theta} = \frac{1}{\sqrt{2}}(\phi_{1} + i \phi_{2}) \Rightarrow  \frac{\overline{\phi}}{\sqrt{2}}(1 + i\delta \theta) = \frac{1}{\sqrt{2}}(\overline{\phi}_{1} + i \delta \phi_{2}) \Rightarrow \delta \theta  = \frac{\delta \phi_{2}}{\overline{\phi}_{1}}\;\; , \;\; \overline{\phi}_{1} = \overline{\phi}   ~.\ee
To obtain the quantum fluctuation of $\phi_{2}$, which will give the fluctuation of the phase, we transform to a canonically normalised field $\chi_{2}$, where 
\be{r15} \chi_{2} = \frac{M_{Pl}}{\sqrt{\xi} \overline{\phi}_{1} } \phi_{2} ~.\ee 
Therefore 
\be{r15a} \delta \theta = \frac{\sqrt{\xi}}{M_{Pl}} \delta \chi_{2}   ~.\ee
Since $\chi_{2}$ is a massless field, it will develop a quantum fluctuation with the standard power spectrum 
\be{r16}  {\cal P}_{\delta \chi_{2}}   = \frac{H^{2}}{ 4 \pi^{2}}    ~.\ee 
The corresponding power spectrum of the phase fluctuations is then 
\be{e17} {\cal P}_{\delta \theta} = \frac{\xi}{M_{Pl}^{2}} {\cal P}_{\delta \chi_{2}} = \frac{\xi H^{2}}{4 \pi^{2} M_{Pl}^{2}}    ~.\ee 
The global $U(1)_{L}$ symmetry ensures that this fluctuation applies at all values of $\theta$. From \eq{e26b} and \eq{n4}, the lepton number and hence baryon number is proportional to $\sin(2 \theta)$, therefore  
\be{e18} \frac{\delta n_{B}}{n_{B}} = \frac{2 \delta \theta}{\tan(2 \theta)}    ~.\ee 
Fluctuations of $\theta$ will therefore produce an isocurvature perturbation  
\be{e19}   {\cal I} = \frac{\Omega_{b}}{\Omega_{dm}} \frac{\delta n_{B}}{n_{B}} = \frac{\Omega_{b}}{\Omega_{dm}}   \frac{2 \delta \theta}{\tan(2 \theta)} ~,\ee 
where we have written this as an equivalent CDM isocurvature perturbation ${\cal I}$ for comparison with the results of Planck \cite{planckiso}. 
Planck gives constraints in terms of the parameter $\beta_{iso}$ for CDM isocurvature perturbations, where 
\be{e20}  \beta_{iso} = \frac{{\cal P}_{I}}{{\cal P}_{R} + {\cal P}_{I} }     ~\ee
and ${\cal P}_{R} \equiv A_{s}$ is the curvature perturbation power. 
From \eq{e19}, 
\be{e21} {\cal P_{I}} =  \left(\frac{\Omega_{b}}{\Omega_{dm}}\right)^{2} 
\frac{4}{\tan^{2}(2 \theta)} {\cal P}_{\delta \theta} = \left(\frac{\Omega_{b}}{\Omega_{dm}}\right)^{2}  
\frac{1}{\tan^{2}(2 \theta)} \frac{\xi H^{2}}{ \pi^{2} M_{Pl}^{2}}  ~.\ee 
The ratio ${\cal P}_{I}/{\cal P}_{R}$ is related to $\beta_{iso}$ by 
\be{r22} \frac{ {\cal P}_{I} }{ {\cal P}_{R}} = \frac{\beta_{iso}}{1 - \beta_{iso} }     ~.\ee
Assuming that $\beta_{iso} \ll 1$, the prediction for $\beta_{iso}$ is 
\be{r23} \beta_{iso} = \frac{\xi H^{2} }{\pi^{2} M_{Pl}^{2} A_{s} \tan^{2}(2 \theta) } \left( \frac{\Omega_{b}}{\Omega_{dm}} \right)^{2}  ~.\ee

In both the metric and the Palatini models, the value of the Einstein frame potential on the plateau during inflation is given by
\be{r25} V_{E}  = \frac{\lambda_{\Phi} M_{Pl}^{4}}{4 \xi^{2}}  ~.\ee
Thus  
\be{r25a} H = \left( \frac{\lambda_{\Phi}}{12} \right)^{1/2} \frac{M_{Pl}}{\xi}   ~\ee
Therefore the model predicts that 
\be{r26} \beta_{iso} = \frac{\lambda_{\Phi}}{12 \pi^{2} \xi A_{s} \tan^{2} (2 \theta) } \left( \frac{\Omega_{b}}{\Omega_{dm}} \right)^{2}  ~.\ee

The value of $\xi$ at $N = 55$ in the metric model is 
$ \xi  = 4.5 \times 10^{4}\, \sqrt{\lambda_{\Phi}}$.
Therefore the metric model predicts that
\be{r29}  \beta_{iso,\, metric} = \frac{3.1 \sqrt{\lambda_{\Phi}} }{\tan^{2} (2 \theta) }  ~.\ee
Planck (2018) obtains the 2-$\sigma$ upper bound $\beta_{iso} < 0.038$ \cite{planckiso}. Therefore the 
metric model satisfies the isocurvature bound if 
\be{r30} \lambda_{\Phi} < 1.5 \times 10^{-4}\, \tan^{4}(2 \theta)  ~.\ee

The value of $\xi$ at $N  = 55$ in the Palatini model is 
$ \xi = 1.2 \times 10^{10}\, \lambda_{\Phi} $. 
Therefore the Palatini model predicts that 
\be{r29}  \beta_{iso,\;Palatini} = \frac{1.2 \times 10^{-5}}{\tan^{2} (2 \theta) }  ~.\ee
Thus the Palatini model easily satisfies the present observational bound independently of $\lambda_{\Phi}$, assuming that $\tan (2 \theta)$ is not unusually small.

Therefore there is an upper bound on $\lambda_{\Phi}$ for the metric model to be consistent with the present bound on isocurvature perturbations. Significantly, baryon isocurvature perturbations close to the present observational limit are possible in the metric model if $\lambda_{\Phi} \sim 10^{-4} \tan^{4}(2 \theta)$. Therefore the metric model would be able to explain a future observation of baryon isocurvature perturbations close to the present limit. In contrast, the Palatini model prediction is much smaller than the present isocurvature bound independently of $\lambda_{\Phi}$, assuming that $\tan(2 \theta)$ is not unusually small.

\section{Quadratic versus Quartic $U(1)_{L}$-Breaking Terms}  

In our analysis we have assumed that leptogenesis is due to the quadratic A-term. In general, we would also expect a quartic C-term to exist. AD leptogenesis via a C-term in non-minimally coupled inflation has been studied in \cite{cline}.
The dynamics in this case is quite different, with the asymmetry being generated during inflation rather than at late times.
Here we consider the limits on $C$ for which the A-term will dominate leptogenesis. The $U(1)_{L}$-breaking potential terms are 
\be{cc1} A (\Phi^{2} + \Phi^{\dagger\,2}) + C (\Phi^{4} + \Phi^{\dagger \, 4})  ~.\ee 
Thus the C-term, for a given $\phi$, will  be less important to the field dynamics than the A-term if $C \phi^{2} \lae A$. We will first consider this condition at $\phi \leq \phi_{*}$, when the potential is $\phi^{2}$ dominated. $C \phi^{2} \lae A$ at $\phi = \phi_{*}$ requires that 
\be{cc2}   C \phi_{*}^{2} = \frac{C m_{\Phi}^{2}}{\lambda_{\Phi}} \lae A \Rightarrow  \frac{C}{\lambda_{\Phi}} \lae \frac{A}{m_{\Phi}^{2}}   ~.\ee
In other words, if the relative contribution of the C-term to the quartic term is no greater than the contribution of the A-term to the quadratic term, the A-term will be dominant at $\phi \leq \phi_{*}$. It is still possible that at $\phi > \phi_{*}$ the quartic term could become important for large enough $\phi$. However, to influence the evolution of the phase and so the baryon asymmetry, $C \phi^{2}$ would also have to become larger than $H^{2}$, where we are considering $C \phi^{2}$ to be approximately the mass squared of the phase field. At $\phi = \phi_{*}$, since we assume that $\phi_{AD} < \phi_{*}$, the phase is not dynamical and so $A \lae H^{2}$ at $\phi = \phi_{*}$. Therefore $C \phi_{*}^{2}$ is also less than $H^{2}$ at $\phi = \phi_{*}$. Then since $H^{2} \propto \phi^{4}$ at $\phi > \phi_{*}$, it follows that $C \phi^{2}$ can never exceed $H^{2}$ at $\phi > \phi_{*}$ if it is less than $H^{2}$ at $\phi = \phi_{*}$. Therefore the condition \eq{cc2} is sufficient for the A-term to dominate the dynamics of the phase field throughout and so to dominate asymmetry generation.  

\section{Conclusions} 

We have discussed a minimal leptogenesis model based on the inflaton mass term Affleck-Dine mechanism introduced in \cite{kls1}. The inflaton sector is a non-minimally coupled complex inflaton, which couples to the Standard Model via RH neutrinos. We have reviewed the analytical predictions for the baryon asymmetry and compared these to some examples of complete numerical solutions, confirming their accuracy. We have also derived conditions for the consistency of the analytical results. Using the analytical expressions, we have shown that the model can easily generate the observed baryon asymmetry. In the case of the model with a metric non-minimally coupled inflaton sector, the model can produce baryon isocurvature perturbations that are close to the present observational bound. The model is also consistent with the range of reheating temperatures, $T_{R} = 10^{6} - 10^{8} \GeV$,  that could be detected in the spectrum of observable primordial gravitational waves predicted by the metric model. This is in contrast to the case of conventional thermal leptogenesis, which requires that $T_{R} \gae 10^{9} \GeV$. The Palatini model predicts that both the baryon isocurvature perturbations and primordial gravitational waves are much smaller than the present and the expected future observational limits.

\section*{Acknowledgements}

KLS was funded by STFC while this project was undertaken.

\end{document}